\documentclass[letterpaper]{mc2025}

\usepackage{todonotes}
\usepackage[caption=false]{subfig}

\usepackage{nomencl} 
\makenomenclature

\title{
Implicit Collision Multiplicity Adjustment for Efficient\\
Monte Carlo Transport Simulation of Reactivity Excursion
}

\addAuthor[variansi@oregonstate.edu]{Ilham Variansyah}{1}
\addAuthor{Ryan G. McClarren}{2}
\addAuthor{Todd S. Palmer}{1}

\addAffiliation{1}{Nuclear Science and Engineering, Oregon State University, Corvallis, Oregon}
\addAffiliation{2}{Aerospace and Mechanical Engineering, University of Notre Dame, Notre Dame, Indiana}

\Abstract{%
We present an implicit collision method with on-the-fly multiplicity adjustment based on the forward weight window methodology for efficient Dynamic Monte Carlo (MC) simulation of reactivity excursion transport problems.
Test problems based on the Dragon experiment of 1945 by Otto Frisch are devised to verify and assess the efficiency of the method.
The test problems exhibit the key features of the Dragon experiment, namely nine orders of magnitude neutron flux bursts followed by significant post-burst delayed neutron effects.
Such an extreme reactivity excursion is particularly challenging and has never been solved with Dynamic MC.
The proposed implicit collision multiplicity adjustment, in conjunction with a simple forced delayed neutron precursor decay technique, profitably trades simulation precision for reduced runtime, leading to an improved figure of merit, enabling efficient Dynamic MC simulation of extreme reactivity excursions.
}


\shortTitle{Monte Carlo implicit collision for reactivity excursion}

\authorHead{Ilham Variansyah, et al.}

\begin{document}

\section{Introduction}\label{sec:intro}
Advances in high-performance computing enable high-fidelity Dynamic Monte Carlo (Dynamic MC) simulations of challenging neutron transport transients~\cite{sjenitzer_dynamic_mc}.
There are several features of time-dependent transport that make Dynamic MC simulations challenging.
In this paper, we are interested in reactivity excursion, which is the key feature in several transport transients of multiplying systems, including reactor criticality accidents and neutron burst experiments.

Reactivity excursion is characterized by rapid growth in the neutron population induced by prompt supercritical reactivity insertion.
Tracking the rapidly growing particle numbers with Dynamic MC is computationally prohibitive.
Furthermore, if the burst caused by the excursion is high enough, significant delayed productions from delayed neutron precursors (DNPs) will follow.
Tracking these delayed productions along with the prompt ones is challenging, particularly due to their significantly different reaction chain time scales, up to the factor of $10^4$~\cite{sjenitzer_dynamic_mc}.

\subsection{The Dragon Experiment} \label{sec:dragon_intro}

\begin{figure}[h!]
    \centering
    \includegraphics[width=0.55\columnwidth]{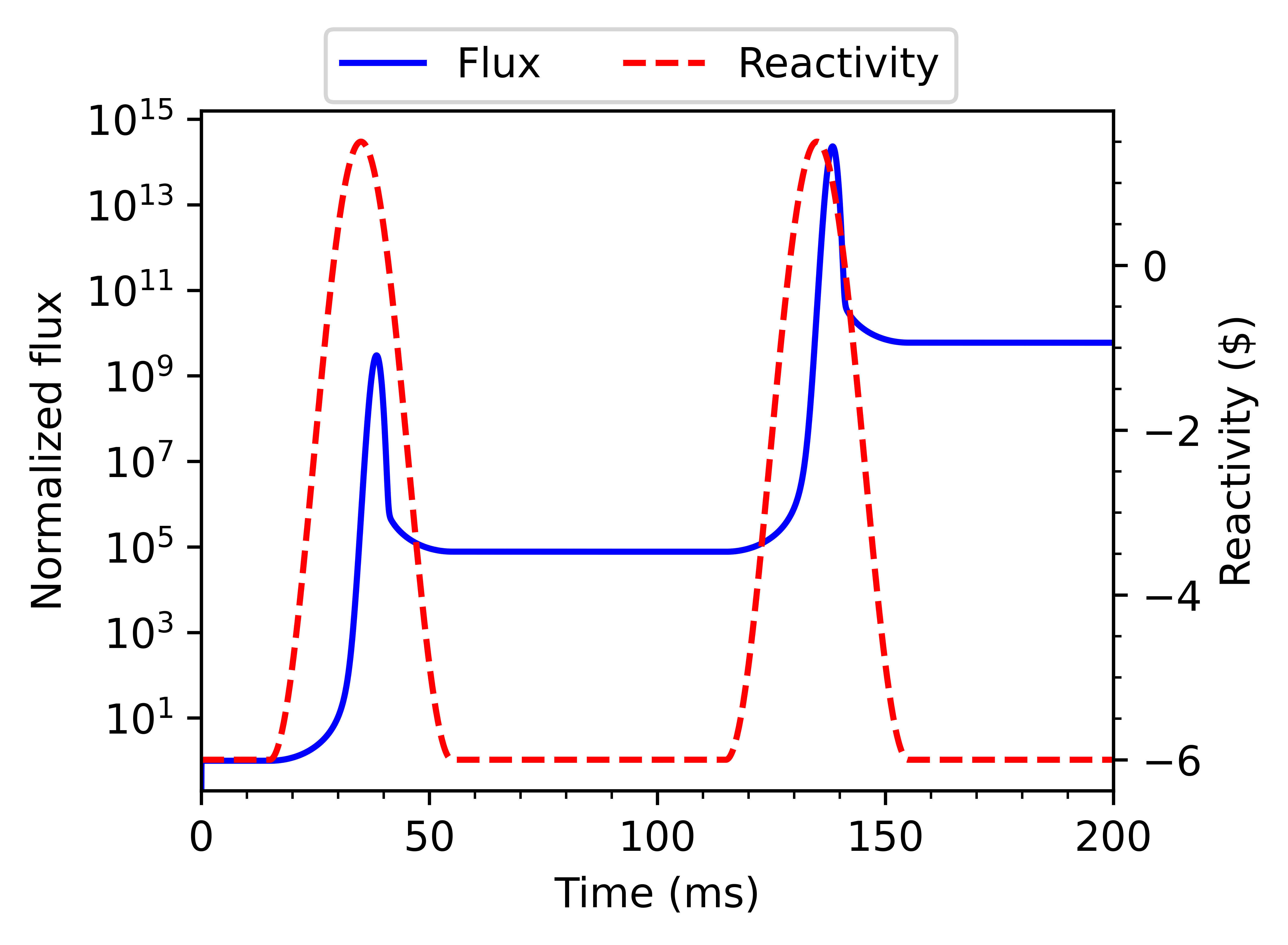}
    \caption{Point kinetic solution of neutron excursions caused by two successive \$1.5 reactivity pulses. The first half of the solution (0 to 100 ms) is similar to the Dragon experiment point kinetic solution by Kimpland et al.~\cite{kimpland2021critical}. This Dragon-like neutron excursion transient is used as the motivational test problem of this paper (referred to as Problem 3, discussed in more detail in~\Cref{sec:dragon}).}
    \label{fig:solution-problem3}
\end{figure}

The Dragon experiment~\cite{frisch1945controlled} of 1945 by Otto Frisch is a good example of how challenging a reactivity excursion transport problem can be.
The experiment involves dropping a high-enriched uranium fuel slug through a hollow fuel block, yielding a long enough prompt-supercritical configuration to cause reactivity excursion.
Kimpland et al.~\cite{kimpland2021critical} recently presented criticality MC calculations of the Dragon assembly configurations. 
By using the resulting effective kinetic parameters, a point-kinetic calculation of the expected power evolution from a single burst was performed.
The resulting point kinetic solution is presented in Figure 12 in~\cite{kimpland2021critical} (but is also illustrated in the first half of \Cref{fig:solution-problem3}).
The reactivity pulse lasts around 40 milliseconds with a peak reactivity of about \$1.5.
This produces an extreme power burst that peaks at about $10^9$ times the initial steady-state power level.
The burst is so strong that the power does not return back to the initial level following the tail of the reactivity pulse.
The power drops quickly for a moment to about $10^{4}$th of the peak level but then stays almost constant (in the milliseconds time scale) due to the delayed neutron productions from the decays of DNPs that are substantially accumulated during the burst.

\Cref{fig:solution-problem3} reproduces and extends the key features from the Kimpland et al. point kinetic solution.
Computational detail on how the point kinetic solution is reproduced and extended is discussed in~\Cref{sec:dragon}.
\Cref{fig:solution-problem3} illustrates what happens if a subsequent burst is produced by re-dropping the high-enriched uranium fuel slug through the hollow fuel block, re-inserting an identical reactivity pulse but with an elevated local initial condition, picking up from the delayed neutron productions from the preceding burst.
This results in a power burst that peaks at about $10^{14}$ times the initial steady-state condition.
These orders of magnitude changes in particle population make such reactivity excursion transients exceptionally challenging for Dynamic MC simulation.

\subsection{Techniques for Dynamic Monte Carlo}

There are techniques for Dynamic MC in the literature that deal with growing (and decaying) neutron populations.
However, their applicability for simulating reactivity excursion can be very limited.

Population Control Techniques (PCTs)~\cite{variansyah_pct} control a time-censused population to a targeted size.
The particle census (the act of stopping and storing particles in a bank whenever they cross the current time boundary) needs to be done frequently enough to limit population growth within each time step.
However, more frequent time census and population control mean more frequent synchronization of particles that adds to runtime and more variances introduced by the sampling processes of the population control~\cite{variansyah_pct}, and in an extreme reactivity excursion transient, the required time census frequency can be prohibitively high.

Branchless collision~\cite{sjenitzer_dynamic_mc} ensures a non-growing population at the expense of decreasing precision in simulation tally results.
The precision loss is due to the highly varying particle weights from the branchless collision sampling. It is expected to be severe for reactivity excursion problems characterized by high frequencies of multiplying collisions.
One way to address precision loss is to pair branchless collision with a weight-based PCT~\cite{variansyah_pct} that respectively splits and roulettes high- and low-weight particles relative to the average weight of the population.
However, this, too, requires sufficiently frequent time censuses to ensure that the precision loss is well-bounded within each time step.

Another technique worth mentioning is weight window~\cite{cooper_ww}.
By using a good estimate of the relative temporal importance of the neutrons, the weight window serves as an effective, on-the-fly, synchronization-free population control.
Nevertheless, the use of temporal weight window is still very limited in the literature~\cite{northrop_vrt}.

The new method proposed in this paper can be seen as a variant of the weight window method.
Fundamentally, the method employs a continuous piecewise temporal weight window and applies the splitting-roulette game based on how long the particle has been traveling in the system.
More specifically, the splitting-roulette game is applied by adjusting the multiplicity of implicit collision, whose secondary particle sampling procedure is similar to the branchless collision, except that none or multiple particles can be produced at the collision.

The rest of the paper is organized as follows.
In \Cref{sec:method}, we first devise motivational problems that exhibit key temporal features of the Dragon experiment.
The problems are simplest in the spatial and energy phase space but challenging in time, which is the focus of this work.
Then, we discuss the proposed implicit collision multiplicity adjustment method in more detail.
The forced DNP decay technique~\cite{sjenitzer_dynamic_mc} is also discussed as it enhances the proposed method to account for the post-burst delayed neutron contribution efficiently.
In \Cref{sec:result}, we present the MC results of the Dragon-like test problems.
Verifications for the Dynamic MC simulations, with and without the proposed method, are performed against highly accurate point kinetic results.
By using the resulting simulation runtime and precisions, the efficiency of the proposed method is characterized and compared against the analog MC results.
Finally, \Cref{sec:summary} summarizes the key findings and discusses future work.

\section{Method} \label{sec:method}

\subsection{Dragon-Like Reactivity Excursion Problems}
\label{sec:dragon}

In this subsection, we devise motivational reactivity excursion problems.
The problems are designed to be simple enough to be accurately solved via point kinetic calculations for verification purposes.
However, the problems need to be challenging enough so that they exhibit key temporal features of the Dragon experiment that stress Dynamic MC simulations.
Furthermore, the problems are designed to be a progressive suite with increasing difficulties, which are needed to characterize and assess the performance of the MC simulations and methods properly.

Let us consider a simple mono-energetic, 0-dimensional (infinite, homogeneous medium), time-dependent fixed-source transport problem:
\begin{equation}
    \frac{1}{v} \frac{d\phi}{dt}
    + \Sigma_t \phi(t)
    =
    (1 - \beta) \nu \Sigma_t \phi(t)
    + \lambda C(t)
    + S,
\end{equation}
\begin{equation}
    \frac{dC}{dt}
    + \lambda C(t)
    =
    \beta \nu \Sigma_t \phi(t),
\end{equation}
with the zero initial condition for the neutron flux and DNP density: $\phi(0) = C(0) = 0$.
We note that the system is a purely fissioning medium, $\Sigma_f = \Sigma_t$, so that the criticality is equivalent to the multiplicity, $k = \nu$, and the reactivity, $\rho(\$)$:
\begin{equation}
    \rho (\$) = \frac{\nu - 1}{\beta \nu}.
    \label{eq:rho}
\end{equation}
The constant source strength $S$ is set to yield a unit prompt neutron flux at the steady-state subcritical condition:
\begin{equation}
    S = [1 - (1 - \beta) \nu] \Sigma_t.
\end{equation}
We drive the transient by introducing two reactivity pulses with identical maximum reactivity $\rho_\text{max} > 0$. Outside the pulses, we have the constant base reactivity $\rho_\text{base} < 0$. The reactivity pulses are realized by making the reaction multiplicity time-dependent:
\begin{equation}
    \nu(t)
    =
    \frac{1}{2}
    \left\{
        \left(
            \nu_\text{max} + \nu_\text{base}
        \right)
        +
        \sin{\left[
            2 \pi \left(
                \frac{t - t_\text{start}}{\Delta t}
                -
                \frac{1}{4}
            \right)
        \right]}
        \left(\nu_\text{max} - \nu_\text{min}\right)
    \right\},
\end{equation}
where $t_\text{start}$ and $\Delta t$ are respectively the start and the width of the reactivity pulses, and $\nu_\text{max}$ and $\nu_\text{base}$ are reaction multiplicities associated with the reactivities $\rho_\text{max}$ and $\rho_\text{base}$ per \Cref{eq:rho}.

By inferring from the point kinetic parameters and solution given in the calculation of the Dragon experiment by Kimpland et al.~\cite{kimpland2021critical}, we set the base reactivity $\rho_\text{base} = -4 \rho_\text{max}$, neutron speed $v$ = 350 km/s, total reaction cross section $\Sigma_t$ = 0.027 /cm, delayed fission production fraction $\beta$ = 0.008, and DNP decay constant $\lambda$ = 0.1 /s.
The first reactivity pulse starts at $t_\text{start}$ = 15 ms and has a width of $\Delta t$ = 40 ms.
The second reactivity pulse is identical to the first one (same $\rho_\text{base}$, $\rho_\text{max}$, and $\Delta t$) but is introduced at $t_\text{start}$ = 115 ms.

\begin{figure}[h!]
    \centering
    \includegraphics[width=0.75\columnwidth]{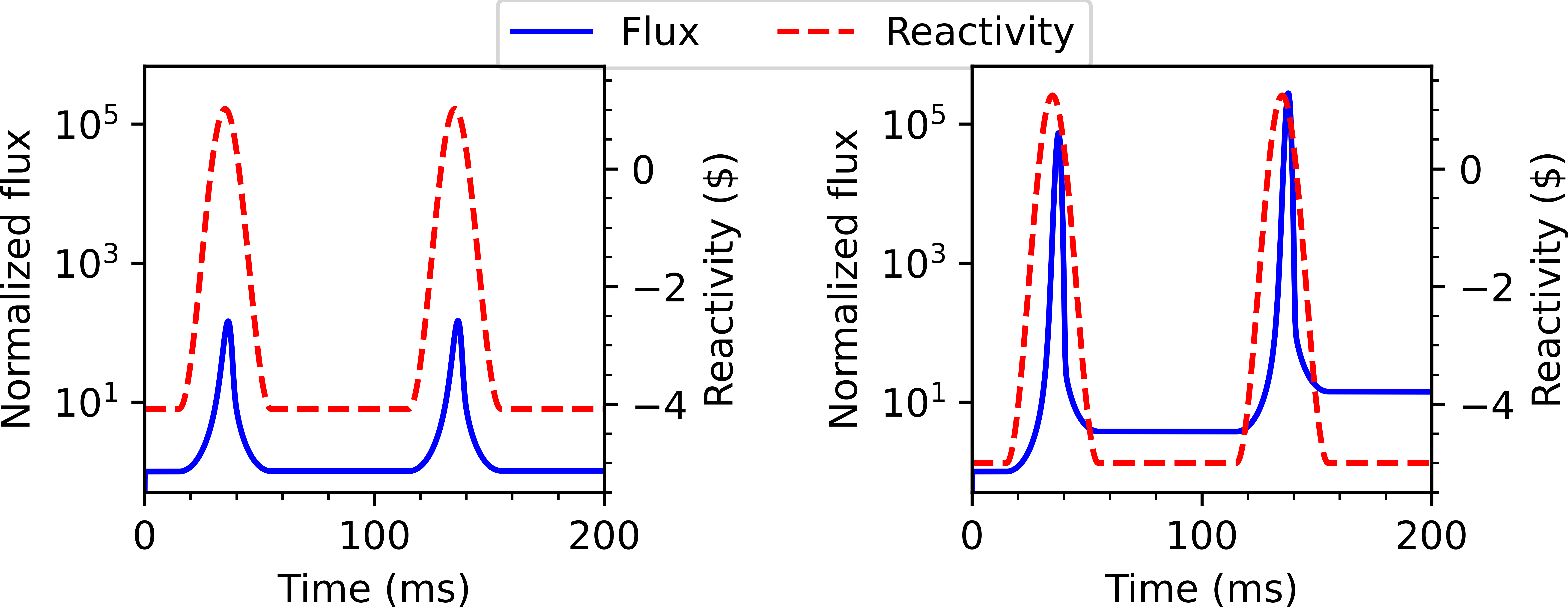}
    \caption{Reference point kinetic solutions of Problem 1 (left, \$1.02 reactivity peaks) and Problem 2 (right, \$1.25 reactivity peaks).}
    \label{fig:solution-problem12}
\end{figure}

We consider three progressive problems with increasing maximum reactivity $\rho_\text{max}$:
\begin{itemize}
    \item Problem 1 ($\rho_\text{max}=\$1.02$): Mild excursions (shown on the left in~\Cref{fig:solution-problem12}),
    \item Problem 2 ($\rho_\text{max}=\$1.25$): Excursions with mild delayed effects (shown on the right in~\Cref{fig:solution-problem12}), and
    \item Problem 3 ($\rho_\text{max}=\$1.5$): Extreme excursions with delayed effects (shown in \Cref{fig:solution-problem3}).
\end{itemize}
The resulting systems of non-homogeneous linear ordinary differential equations are solved brute-forcefully using Scipy's initial value problem solver \texttt{integrate.solve\_ivp}.
The default Explicit Runge-Kutta method of order 5(4), a.k.a. RK45, is used with a maximum allowed time step of size $10^{-4}$ ms, which is about one-tenth of the neutron mean free time $(v \Sigma_t)^{-1}$, and relative and absolute tolerances of $10^{-9}$ and $10^{-18}$.
The calculation took less than two seconds on an Apple M2 chip for the hardest Problem 3.

\Cref{fig:solution-problem3} shows the solution for Problem 3.
The first half of the solution (0-100 ms) is similar to the solution of the Dragon experiment by Kimpland et al. (Figure 12 in~\cite{kimpland2021critical}).
In the Kimpland et al. study, the neutron burst solution is only shown for $t \in $ 700-800 ms, during which the reactivity pulse is introduced.
The solution in $t<$ 700 ms is not shown as it is essentially flat.
This is due to the small neutron mean free time (about $10^{-3}$ ms) that quickly evolves the zero flux initial condition to the prompt steady-state subcritical condition, within 1\% in less than 0.1 ms, which is almost like a step function in the hundreds of ms time frame.
For the Dragon-like problems we devised, we cut that first 700 ms for efficiency.
Discussions on other key features of Problem 3 can be found in \Cref{sec:dragon_intro}.

\Cref{fig:solution-problem12} shows the solutions for Problems 1 and 2. In Problem 1, a mild \$1.02 reactivity pulse causes more than a hundred-fold increase in the neutron flux.
However, different from Problem 3, after each neutron burst, the neutron flux decays back down to the pre-burst level, which is essentially driven by the fixed source.
This is because the neutron bursts are not high enough to yield post-burst delayed neutron production that is stronger than the fixed source strength.
In Problem 2, on the other hand, the \$1.25 reactivity pulse causes a high enough burst for us to observe the significant effect of the post-burst delayed neutron production.
Later in \Cref{sec:result}, we will find that these two problems can be reasonably solved with the analog MC simulation.
On the other hand, Problem 3 is prohibitively expensive, but not if the proposed method is applied.

\subsection{Branchless and Implicit Collisions}

Branchless collision~\cite{sjenitzer_dynamic_mc} applies a combined effect of all possible reactions to the colliding particle by (1) modifying the particle weight based on the total production cross-section $\nu_t\Sigma_t$:
\begin{equation}
    w'
    =
    w
    \frac{\nu_t\Sigma_t}
    {\Sigma_t},
    \label{eq:branchless}
\end{equation}
and (2) changing the colliding particle phase space as if it is a secondary particle coming out from one of the possible reactions, with probabilities proportional to the associated production cross-section.
This results in a non-branching process: one particle in, one particle out.

The total production cross-section includes productions from fission ($\nu\Sigma_f$), scattering ($\nu_s\Sigma_s$), and other reactions.
Essentially, this also includes ``productions" from non-multiplying reactions like radiative capture ($\nu_\gamma\Sigma_\gamma$), even though they have a multiplicity of zero ($\nu_\gamma=0$).
From this observation, one can see that branchless collision is an extension of the more well-known MC transport technique implicit capture (a.k.a survival biasing) as it implicitly applies all possible reactions, not just capture, to the colliding particle.
In other words, branchless collision is an implicit collision---more specifically, it is an implicit collision with unit multiplicity. In the next subsection, we discuss the general form of implicit collision and the proposed on-the-fly multiplicity adjustment.

%

\subsection{Implicit Collision with On-The-Fly Multiplicity Adjustment}

As discussed in the previous section, branchless collision is an implicit collision with a unit multiplicity.
If we denote the multiplicity of implicit collision as $\hat{\nu}$, then for branchless collision $\hat{\nu}=1$.
However, in general, the multiplicity of an implicit collision can be any positive real number, $\hat{\nu}>0$.

Given a multiplicity $\hat{\nu}$, the number of secondary particle productions can be sampled as $\left\lfloor \hat{\nu} + \xi \right\rfloor$, where $\xi$ is a uniform random number from zero to one.
The phase space of each secondary particle is sampled as if it is coming out from one of the possible reactions, with probabilities proportional to the associated production cross-section.
But more importantly, to maintain a non-biased MC simulation, each of the produced secondary particles is assigned a uniform weight of
\begin{equation}
    w'
    =
    \frac{1}{\hat{\nu}}
    w'_\text{total}
    ,
    \quad\quad
    w'_\text{total}
    =
    w
    \frac{\nu_t\Sigma_t}
    {\Sigma_t}.
    \label{eq:implicit_coll}
\end{equation}
Note that for branchless collision, the entire total post-collision weight $w'_\text{total}$ is carried by the single secondary particle, while in general, it is uniformly distributed among the $\hat{\nu}$ (on average) particles.

The proposed method is based on the forward-based weight window methodology for global MC transport calculation~\cite{cooper_ww}.
By using an estimate of the transport solution, one can evenly distribute the MC particles throughout the problem domain.
This is especially useful for reactivity excursion problems where the physical particles are highly concentrated in the time domain.
In this work, we perform the weight window-based splitting-roulette game whenever a collision occurs.
However, rather than directly splitting or rouletting the particle, the splitting-roulette game is effectively embedded during the sampling of the number of secondary particle productions.

Given an estimate of the transport solution $\hat{\phi}(t)$, we determine the target secondary particle weight as follows:
\begin{equation} 
    w'
    =
    w
    \frac
        {\hat{\phi}(t_\text{coll})}
        {\hat{\phi}(t_\text{birth})},
    \label{eq:weight_secondary}
\end{equation}
where $t_\text{coll}$ and $t_\text{birth}$ are, respectively, the time of collision and time of birth (either from fixed source or secondary production) of the colliding particle, streaming through the medium with the starting weight $w$.
In this work, we employ a continuous, piecewise linear $\hat{\phi}(t)$ obtained from a point kinetic calculation on uniform time bins.
Finally, given the target weight $w'$, we can determine the implicit collision multiplicity $\hat{\nu}$ based on \Cref{eq:implicit_coll}:
\begin{equation}
    \hat{\nu}
    =
    \frac{w'_\text{total}}
    {w'}
    =
    \frac{\hat{\phi}(t_\text{birth})}
    {\hat{\phi}(t_\text{coll})}
    \left(
        \frac{\nu_t\Sigma_t}
        {\Sigma_t}
    \right).
    \label{eq:multiplicity}
\end{equation}
\Cref{eq:multiplicity} suggests that when a particle moves and collides at a time at which the particle density is estimated to be higher, $\hat{\phi}(t_\text{coll}) > \hat{\phi}(t_\text{birth})$, the implicit collision multiplicity $\hat{\nu}$ decreases and the weight of the secondary particles $w'$ increases, per \Cref{eq:weight_secondary}.
On the other hand, when a particle moves and collides at a time at which the particle density is estimated to be lower, $\hat{\phi}(t_\text{coll}) < \hat{\phi}(t_\text{birth})$, the implicit collision multiplicity $\hat{\nu}$ increases and the weight of the secondary particles $w'$ decreases.

\subsection{Forced DNP Decay}

If the reactivity excursion is strong enough, the neutron burst is shortly followed by a significant delayed neutron effect.
This is evident in the case of Problems 2 and 3.
There are challenges in simulating this post-burst delayed effect.
Only a small fraction of the delayed neutrons produced land within the short time interval of interest.
With the moderate decay constant of $\lambda=0.1$ /s, less than 1\% of the delayed neutrons are emitted in the next 100 ms.
Most of the delayed neutrons escape the domain of interest.
This issue is actually compounded by the proposed implicit collision multiplicity adjustment.
This is because the significant delayed neutrons, including those escaping the domain, have very high weights, as they are produced during the reactivity burst, during which the weight roulette effect is dominant (reduced $\hat{\nu}$ and increased $w'$).

To address this issue, we employ a simple version of the forced DNP decay technique~\cite{sjenitzer_dynamic_mc}.
When a delayed neutron is sampled, we force it to be emitted within the rest of the problem domain.
This is achieved by replacing the exponential analog probability density function of the emission time:
\begin{equation}
    P_d(t)
    = 
    \lambda
    e^{-\lambda (t - t_\text{coll})},
    \quad\quad
    t_\text{coll}<t,
\end{equation}
with a uniform one:
\begin{equation}
    P_d^*(t)
    = 
    \frac{1}
    {t_\text{boundary} - t_\text{coll}},
    \quad\quad
    t_\text{coll}<t<t_\text{boundary},
    \label{eq:pdf_forced}
\end{equation}
where $t_\text{boundary}$ is the time boundary of the domain.
To maintain a non-biased MC simulation, the forced delayed neutron is assigned with the weight
\begin{equation}
    w_d^*
    =
    w'
    \frac{P_d(t_\text{decay})}
    {P_d^*(t_\text{decay})},
\end{equation}
where $t_\text{decay}$ is the emission time sampled from the uniform distribution and $w'$ is the prompt secondary neutron weight per \Cref{eq:weight_secondary}.

Note that the delayed neutron essentially moves in time from $t_\text{coll}$ to $t_\text{decay}$, at which the estimated neutron flux, or the weight window, may be different---i.e., $\hat{\phi}(t_\text{decay}) \neq \hat{\phi}(t_\text{coll})$.
Therefore, for consistency, we apply the splitting-roulette game to the delayed neutron.
The target weight is now
\begin{equation} 
    w'_d
    =
    w'
    \frac
        {\hat{\phi}(t_\text{decay})}
        {\hat{\phi}(t_\text{coll})},
\end{equation}
and we split-roulette the particle with multiplicity
\begin{equation}
    \hat{\nu}_d
    =
    \frac{w_d^*}
    {w'_d}
    =
    \frac{\hat{\phi}(t_\text{coll})}
    {\hat{\phi}(t_\text{decay})}
    \frac{P_d(t_\text{decay})}
    {P_d^*(t_\text{decay})}.
    \label{eq:multiplicity}
\end{equation}
In other words, $\lfloor \hat{\nu}_d + \xi \rfloor$ particles, each with the weight $w'_d$, are produced instead.

\section{Results and Discussions} \label{sec:result}

The proposed implicit collision with multiplicity adjustment and forced DNP decay is implemented to MC/DC~\cite{variansyah_mcdc}, a Python-based MC code with just-in-time compilations.
The three progressive Dragon-like reactivity excursion problems devised in \Cref{sec:dragon} are used for verification and to assess the efficiency of the proposed method.

We run MC simulations of the three problems with analog, branchless, and the proposed implicit collisions.
Quantities of interest (QoIs) are the time-averaged neutron fluxes in uniform bins of size 2 ms (100 total tally bins for each problem).
Branchless collision is run as an implicit collision with a fixed unit multiplicity, while the implicit collision referred to in the rest of this paper uses the proposed on-the-fly multiplicity adjustment.
The forced DNP decay technique is employed for both branchless and implicit collision.
Piecewise linear estimates of the flux solutions obtained from point kinetic calculations with 100 uniform bins (same as the QoIs) are used for the multiplicity adjustment as the $\hat{\phi}(t)$.
The sensitivity of the implicit collision multiplicity adjustment to the fidelity of $\hat{\phi}(t)$ is also assessed.
All of the MC simulations were run with distributed memory parallelization on the Lawrence Livermore National Laboratory compute platform Dane, having Intel Sapphire Rapids CPU architecture with 112 cores per node. Ten compute nodes were used, giving a total of 1120 CPU cores.

\subsection{Verification}

\begin{figure}[h!]
    \centering
    \includegraphics[width=1.0\columnwidth]{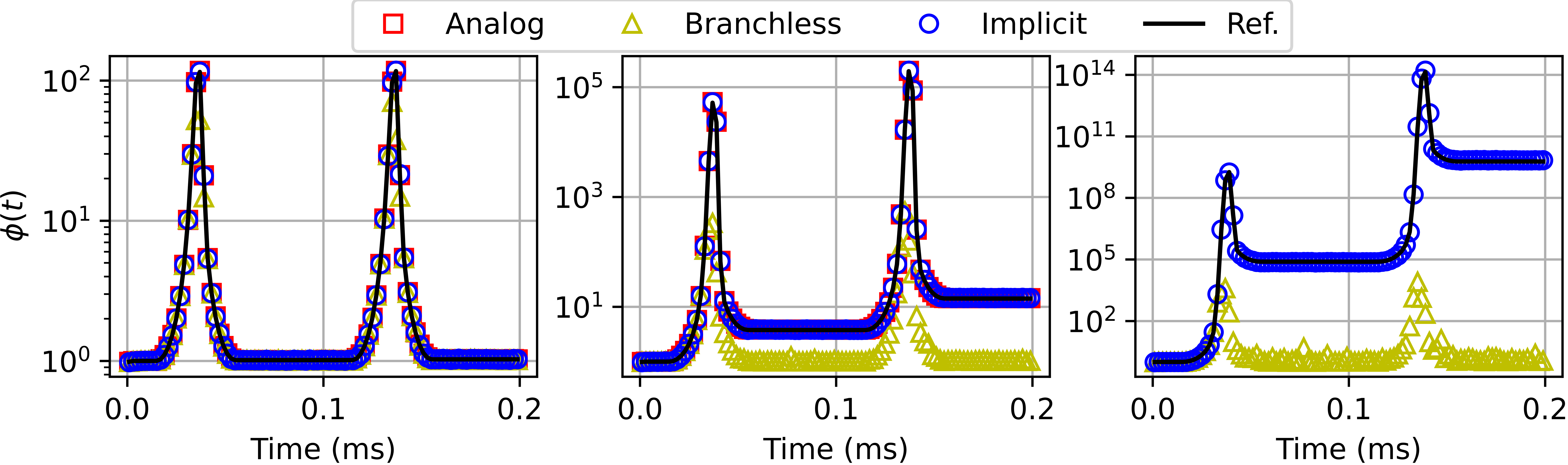}
    \caption{MC solutions of Problems 1, 2, and 3 (left to right) for analog, branchless, and multiplicity-adjusted implicit collisions, all run with $10^7$ source particle histories.}
    \label{fig:MC-solution}
\end{figure}

\Cref{fig:MC-solution} shows the MC solutions (mean values) for Problems 1, 2, and 3 run with $10^7$ source particle histories.
Both analog and implicit collisions match the reference point kinetic solutions very well for Problems 1 and 2.
The most challenging problem, Problem 3, is prohibitively expensive to solve with analog collision; only implicit collision can generate a solution that well matches the reference solution.
Branchless collision poorly matches the reference solution for all problems.
Particularly, it fails to get the excursion peaks, resulting in the subsequent failure to capture the post-burst delayed effects (more on this in the next subsection).

\begin{figure}[h!]
    \centering
    \includegraphics[width=1.0\columnwidth]{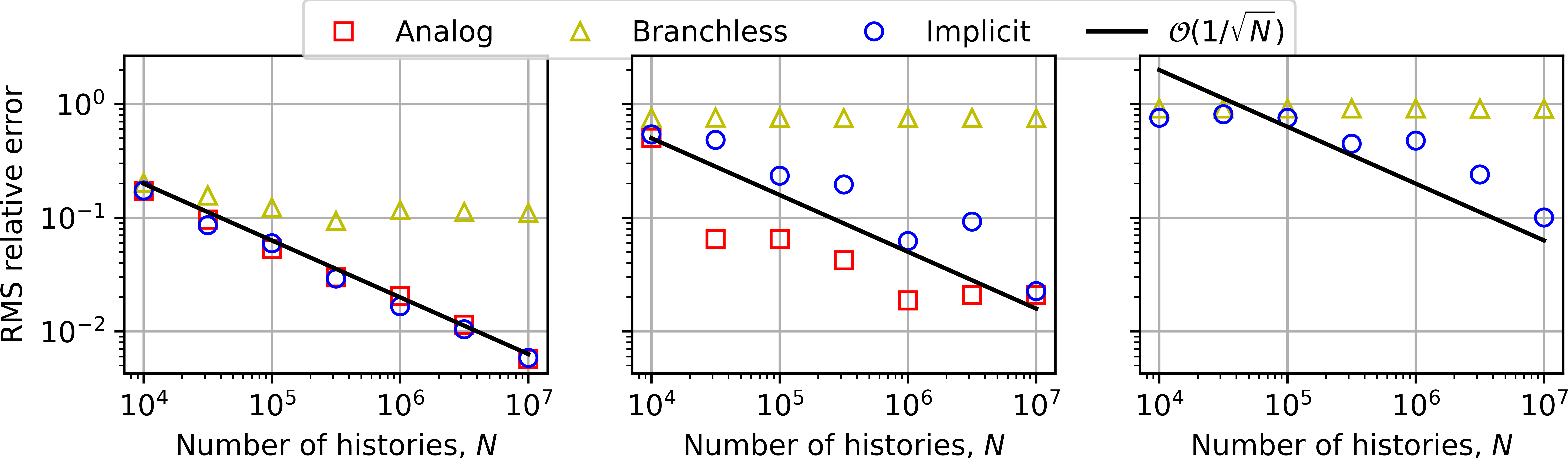}
    \caption{Convergences of the MC solution root-mean-square relative error of Problems 1, 2, and 3 (left to right) for analog, branchless, and multiplicity-adjusted implicit collisions.}\label{fig:MC-convergence}
\end{figure}

\Cref{fig:MC-convergence} shows the convergences of the MC solution root-mean-square (RMS) of the relative errors over the 100 tally bins for Problems 1, 2, and 3.
The $\mathcal{O}(1/\sqrt{N})$ convergence rate, where $N$ is the number of source particle histories, is not only evident in the RMS of the relative standard deviations (not shown here) but also in the RMS of the relative error of the analog and implicit collision for the higher values of $N$.
This verifies the analog MC simulation and the proposed implicit collision with multiplicity adjustment and forced DNP decay.
It is worth noting that the magnitude of the error increases with the difficulty of the problem, and the error of the implicit collision is similar to the analog one for Problem 1, but it is higher for the harder Problem 2.

\subsection{Efficiency}

\begin{figure}[h!]
    \centering
    \includegraphics[width=0.9\columnwidth]{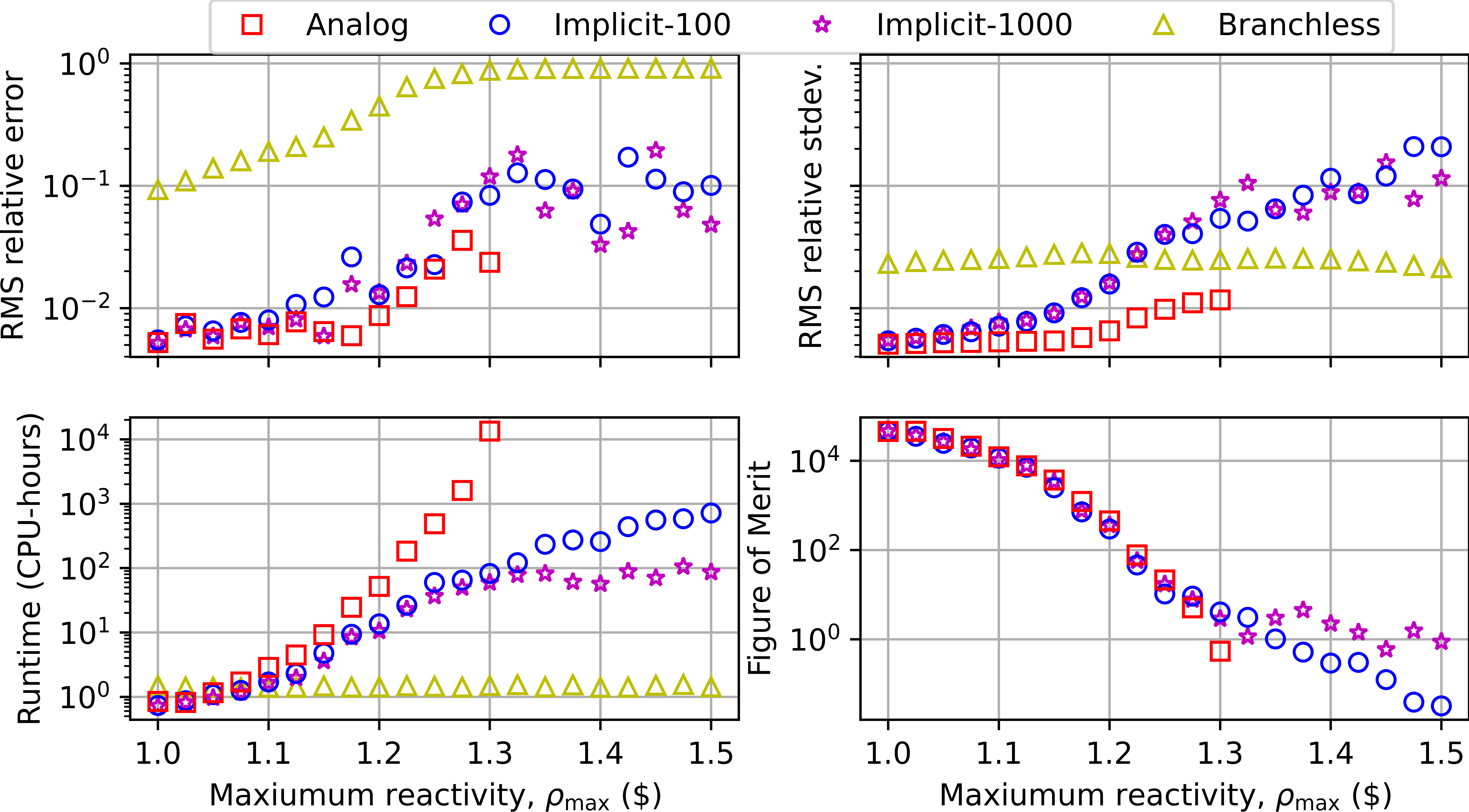}
    \caption{Trends of the MC solution root-mean-square relative errors and standard deviations (top), as well as simulation runtimes (bottom-left) and figures of merit (bottom-right) for analog, branchless, and multiplicity-adjusted implicit collisions (with different numbers of neuron flux estimate bins).}
    \label{fig:MC-efficiency}
\end{figure}

To assess the efficiencies of the methods, we use the Figure of Merit (FOM):
\begin{equation}
    \text{FOM}
    =
    \frac{1}
    {\text{Runtime} \times \text{RMS}(\sigma_r)^2},
\end{equation}
where $\text{RMS}(\sigma_r)$ is the RMS of the relative standard deviations over the 100 tally bins.
\Cref{fig:MC-efficiency} shows the RMS of the relative errors, RMS of the relative standard deviations, runtime, and the FOM for several Dragon-like problems with increasing maximum reactivity $\rho_\text{max}$.
The runtime is presented in the unit of CPU-hours, the product of the number of parallel CPU cores used and the actually recorded runtime.
All of the simulations are run with $N=10^7$, at which the $\mathcal{O}(1/\sqrt{N})$ convergence rates are already evident for the analog and implicit collisions, even for the hardest problem ($\rho_\text{max}=\$1.5$) as shown in \Cref{fig:MC-convergence}.

For the analog and implicit collisions, their RMS of the relative errors are within the expected ranges based on their respective RMS of the relative standard deviations for all $\rho_\text{max}$ values (see the two top figures in \Cref{fig:MC-efficiency}).
This indicates that the asymptotic stochastic behavior of the MC simulation has been achieved with the given number of histories and derived analyses, such as the calculation of FOM, is appropriate.

This is not evident, however, for the branchless collision.
Its RMS of relative errors and standard deviations indicate that the method is precisely wrong, with the given number of histories, for all $\rho_\text{max}$ values.
Note that branchless collision simulates all the problems with identical, non-branching, random walk tracks.
This can be observed from its constant runtime across all $\rho_\text{max}$ values (bottom-left of \Cref{fig:MC-efficiency}); its runtime for the hardest problem is the same as the easiest one.
Branchless collision excessively trades precision for decreased runtime.
Significantly high numbers of particle histories are needed for branchless collision to get appropriate results, especially for the harder excursion problems.

The implicit collision performs similarly to the analog collision for mild reactivity excursion (low $\rho_\text{max}$) problems (see all figures of \Cref{fig:MC-efficiency}).
However, as the excursion gets stronger ($\rho_\text{max}$ increases), analog collision yields more precise simulations (top-right of \Cref{fig:MC-efficiency}).
Nevertheless, the analog collision runs significantly longer, and the rate at which its runtime increases grows with increasing $\rho_\text{max}$ (bottom-left of \Cref{fig:MC-efficiency}), which eventually makes analog collision prohibitively expensive.
Implicit collision profitably trades precision for decreased runtime, leading to a more maintained FOM.
These make FOM of the analog collision decreases more rapidly as $\rho_\text{max}$ increases, making the proposed implicit collision a superior method for extreme excursion problems.

Finally, we run the same efficiency assessment for implicit collision with a higher fidelity of the flux solution estimate  $\hat{\phi}(t)$ for the multiplicity adjustment, where 1000 uniform tally bins are used, instead of the original 100.
It is found that a more accurate estimate of the flux solution helps in reducing simulation runtime while keeping the simulation precision, leading to improved FOM for extreme reactivity excursion problems.

\section{Conclusions and Future Work} \label{sec:summary}

We present an implicit collision method with on-the-fly multiplicity adjustment based on the forward weight window methodology.
Test problems based on the Dragon experiment of 1945 by Otto Frisch are devised to verify and assess the efficiency of the method.
The test problems exhibit the key features of the Dragon experiment, namely nine orders of magnitude neutron flux bursts followed by significant post-burst delayed neutron effects.
Such an extreme reactivity excursion is particularly challenging and has never been solved with Dynamic MC.
The proposed implicit collision multiplicity adjustment, in conjunction with a simple forced delayed neutron precursor decay technique, profitably trades simulation precision for reduced runtime, leading to an improved figure of merit, enabling efficient Dynamic MC simulation of extreme reactivity excursions.

Future work includes (1) implementing the method for Dynamic MC calculation of the real Dragon experiment; (2) investigating how the method affects the scalability of parallel simulations; (3) investigating how the method interacts with other MC techniques when used in conjunction; and (4) exploring the method applications on other reactivity excursion problems, including nuclear reactor criticality accidents and neutron burst experiments.

\section*{Acknowledgements}
This work was supported by the Center for Exascale Monte-Carlo Neutron Transport (CEMeNT), a PSAAP-III project funded by the Department of Energy, grant number: DE-NA003967.

\bibliographystyle{mc2025}
\bibliography{main}

\end{document}